# Color simulation for multilayered thin films using Python

Dongik Lee and Seunghun Lee[†]

Department of Physics, Pukyong National University, Busan 48513, Republic of Korea

[†]seunghun@pknu.ac.kr

## Abstracts

Physical insight into a material can be first gained by its color since the reflectance spectrum from an object reflects its microstructure and complex reflective indices. We here present a comprehensive overview of electrodynamics and optics related to reflectance spectra and color and provide an open-source Python code for simulating reflectance spectra and extracting color values. The validity and applicability of the code are demonstrated through comparative analysis with both literature and experimental data.

## Introduction

The color of an object is one of the foremost characteristics that captures our attention. In the 17th and 18th centuries, which marked the Age of Enlightenment, vigorous effort was made to understand natural phenomena, and especially the study of light and color was systematized by luminaries such as Newton and Fresnel during this era. The simulation of color based on electrodynamics and optics became instrumental in predicting and elucidating experimental results in condensed matter physics and materials science. Nowadays, numerous software and browsers can be found for color simulation, but it is difficult to customize and adjust parameters according to personal demands.

Recently, many individuals have used programming languages for dealing with their own tasks, such as data analysis, fitting, as well as simulation. Among programming languages, Python stands out because of its accessibility, user-friendly program, and the availability of diverse libraries encompassing essential functions for simulation, including calculation, visualization, and interpolation.

Here, we share a Python code for simulating the color of thin-film multilayers while delivering ample background on the underlying physics and coding – from basic electrodynamics, including the Fresnel equations, to the conversion of reflectance spectrum to RGB values based on color matching function. The code requires only information on the complex refractive indices (for visible range) and thickness of each layer and a specific angle of incidence. To validate the code, we compare the simulated color using our Python code with both literature for $SiO_2$ film on a Si substrate and our experimental results for $SnO_2$ thin film with a thickness gradient on a Si substrate. The code is straightforward and allows individuals with some Python background knowledge to easily modify or add conditions, facilitating the color simulation for various conditions and incorporation with Python codes for other purposes. We hope that this code helps students and researchers working on materials science and AI-based research in various shapes.

# Theoretical background

*Reflection from a bulk surface*

When light is incident on an object, it undergoes multiple absorptions, reflections, and transmission at the surface, and the color of an object is determined by the spectrum of reflected light from the surfaces. Fresnel equations describe the reflected and transmitted amplitudes with respect to that of the incident wave, and it can be derived from the boundary conditions of electric and magnetic fields at an interface.

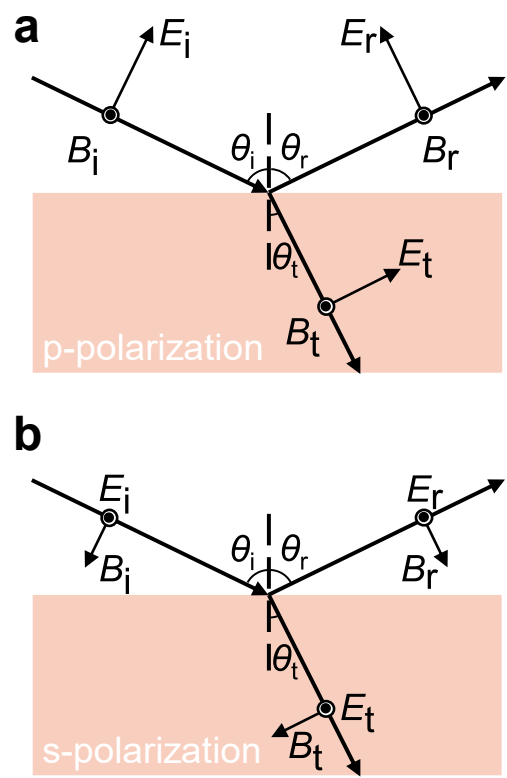

**Figure 1.** Wave incidence at angle $\theta_i$ onto an interface between two media with refractive indices $n_1$ and $n_2$: (a) electric field (**E**) parallel to the plane of incidence (p-polarization) and (b) **E** perpendicular to the plane of incidence (s-polarization). **E** and magnetic field (**B**) are mutually perpendicular (*i.e.*, propagation vector **k**=**E**×**B**). The subscripts i, r, and t represent incident, reflected, and transmitted, respectively.

We define the plane of incidence that contains the surface normal vector and the propagation vector of incident light and determine p or s-polarization from the direction of **E** oscillation with respect to the plane of incidence (Figure 2). The p-polarization indicates that **E** oscillates parallel to the plane of incidence. Here, we first derive the Fresnel equations for the case of polarization parallel to the plane of incidence (*i.e.*, p-polarization). Since the parallel component of **E** and the normal component of **B** with respect to the interface need to be continuous, we arrive:

$$E_i \cos\theta_i - E_r \cos\theta_r = E_t \cos\theta_t \tag{1.1}$$

$$B_i + B_r = B_t \tag{1.2}$$

where the subscripts i, r, and t represent the incident, reflected, and transmitted, respectively. Since $B = E \cdot \frac{n}{c}$ ($n$: refractive index, $n \equiv \frac{c}{v}$) where $v$ is the speed of an electromagnetic wave in matter[1], Eq. 1.2 says:

$$n_i(E_i + E_r) = n_t E_t \tag{2}$$

where $n_i$ and $n_t$ are the refractive indices of media before and after transmission. Given the laws of reflection ($\theta_r = \theta_i$) and refraction (i.e., Snell's law; $n_i \sin\theta_i = n_t \sin\theta_t$) and solving the Eq. 1.1 and 1.2, we obtain the reflection and transmission coefficients for p-polarized light:

$$r_p \equiv \frac{E_r}{E_i} = \frac{n_t \cos\theta_i - n_i \cos\theta_t}{n_t \cos\theta_i + n_i \cos\theta_t} \tag{3.1}$$

$$t_p \equiv \frac{E_t}{E_i} = \frac{2n_i \cos\theta_i}{n_t \cos\theta_i + n_i \cos\theta_t} \tag{3.2}$$

For s-polarization (Fig. 2b), the boundary conditions for **E** and **B** become:

$$E_i + E_r = E_t \tag{4.1}$$

$$-B_i \cos\theta_i - B_r \cos\theta_r = -B_t \cos\theta_t \tag{4.2}$$

The reflection and transmission coefficient for s-polarized light can be obtained by the same procedure.

$$r_s \equiv \frac{E_r}{E_i} = \frac{n_i \cos\theta_i - n_t \cos\theta_t}{n_i \cos\theta_i + n_t \cos\theta_t} \tag{5.1}$$

$$t_s \equiv \frac{E_t}{E_i} = \frac{2n_i \cos\theta_i}{n_i \cos\theta_i + n_t \cos\theta_t} \tag{5.2}$$

The reflectance ($R$) and transmittance ($T$) are defined by the ratio of reflected and transmitted intensity (average power per unit area) to the incident intensity ($I = \frac{1}{2}\epsilon v E^2$), thus:

$$R \equiv \frac{I_r}{I_i} = \left|\frac{E_r}{E_i}\right|^2 = |r|^2 \tag{6.1}$$

$$T \equiv \frac{I_t \cos\theta_t}{I_i \cos\theta_i} = \left(\frac{n_t \cos\theta_t}{n_i \cos\theta_i}\right)\left|\frac{E_t}{E_i}\right|^2 = \left(\frac{n_t \cos\theta_t}{n_i \cos\theta_i}\right)|t|^2 \qquad (6.2)$$

The reflectance depends on the polarization of light, and we consider the natural light or unpolarized light in this. The reflectance of unpolarized light ($R_n$) can be given by the averaged reflectance for all incident waves with different polarization angles:

$$R_n = \frac{1}{2\pi}\int_{-\pi}^{\pi} R_\psi \, d\psi = \frac{(R_p + R_s)}{2} \qquad (7)$$

where $\psi$ is the polarization angle[2].

*Reflection from a (multilayered) thin film*

For a thin film or multilayered structure, we need to consider absorption and optical interference due to multiple reflections, transmissions, and absorptions at interfaces underneath the surface.

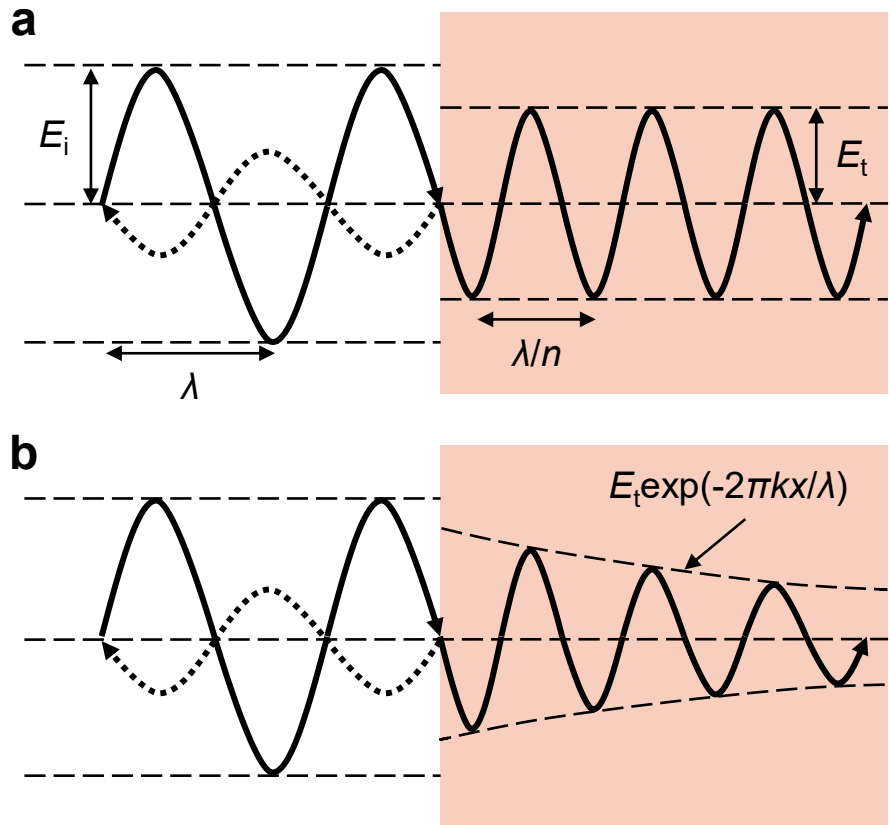

**Figure 2.** Wave propagation in a medium with (a) zero and (b) nonzero extinction coefficient ($k$).

The absorption can be formulated by incorporating a complex refractive index into the wave function. The wave function of a plane wave traveling in the x direction can be written in complex notation:

$$\tilde{\mathbf{E}} = \tilde{\mathbf{E}}_0 \exp[i(\omega t - \tilde{K}x)] \tag{8}$$

where $\tilde{\mathbf{E}}_0$, $\omega$, $t$, $\tilde{K}$, and $x$ are the complex wave amplitude, angular frequency, time, complex wavenumber, and position, respectively. The complex wavenumber can be written as:

$$\tilde{K} \equiv \frac{2\pi \tilde{N}}{\lambda} \tag{9}$$

where $N$ is the complex refractive index. $N$ is defined as:

$$\tilde{N} \equiv n - ik \tag{10}$$

where $n$ and $k$ are the refractive index and extinction coefficient, respectively. By substituting Eq. 9 and Eq. 10 into Eq. 8, we obtain the electromagnetic wave expression taking attenuation (*i.e.*, exponential decay) due to the absorption into account as:

$$\tilde{\mathbf{E}} = \tilde{\mathbf{E}}_0 \exp\left(-\frac{2\pi k x}{\lambda}\right) \cdot \exp\left[i\left(\omega t - \frac{2\pi n}{\lambda}x\right)\right]. \tag{11}$$

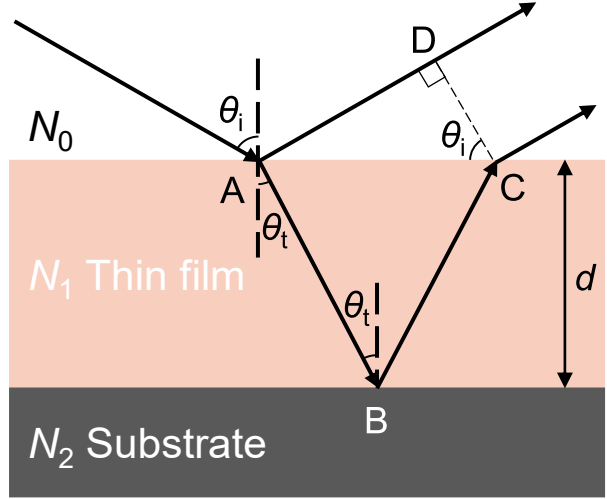

**Figure. 3.** Optical interference in a thin film. The interference between the two reflected lights from both upper and lower boundaries needs to be taken into account. $\theta_i$ and $\theta_t$ are the angle of incidence and refraction, respectively. $\theta_i$ and $\theta_t$ are equal to the $\angle ACD$ and half of the $\angle ABC$, respectively, according to the reflection law. $d$ is the thickness of the film.

In addition to absorption, we need to consider optical interference. Figure 4 shows the presence of two reflected lights from the interfaces between 1) air and thin film (upper) and 2) thin film and substrate (lower). The two reflected light waves will interfere either constructively or destructively, depending on the phase difference between the two waves. The path length difference between the two waves is $\overline{AB} + \overline{BC} - \overline{AD}$. The paths $\overline{AB}$ and $\overline{BC}$ traverse a medium with a refractive index of $N_1$, while path $\overline{AD}$ does a medium with a refractive index of $N_0$. Therefore, the phase difference ($\alpha$) is obtained by applying the different wavenumbers ($K = \frac{2\pi N}{\lambda}$) as:

$$\alpha = \frac{2\pi N_1}{\lambda}(\overline{AB} + \overline{BC}) - \frac{2\pi N_0}{\lambda}\overline{AD}. \tag{12}$$

Considering the incident and transmitted angles, we have $\overline{AD} = \overline{AC}\sin\theta_i$ and $\overline{AC} = 2d\tan\theta_t$, which leads to

$$\overline{AD} = 2d\tan\theta_t \cdot \sin\theta_i. \tag{13}$$

By incorporating Snell's law ($\sin\theta_i = \sin\theta_t N_1/N_0$),

$$\overline{AD} = 2d\frac{\sin^2\theta_t}{\cos\theta_t} \cdot \frac{N_1}{N_0}. \tag{14}$$

Using Eq. 14 and $\overline{AB} = \overline{BC} = d/\cos\theta_t$, we arrive:

$$\alpha = \frac{4\pi d N_1}{\lambda} \cdot \cos\theta_t. \tag{15}$$

The plane wave considering the phase difference becomes:

$$\exp[i\{\omega t - (Kr + \alpha)\}] = \exp[i(\omega t - Kr)] \cdot \exp(-i\alpha). \tag{16}$$

Eq. 16 indicates that the phase difference can be imposed by multiplying $\exp(-i\alpha)$. To consider the optical interference, we need to take the superposition of light waves, which are reflected from both upper and lower boundaries. The primary wave (*i.e.*, directly reflected from the film surface) becomes $E_1 = r_{01} E_i$, in which $r_{01}$ is the reflection coefficient determined by the Fresnel equations for the boundary between the air and thin film. The secondary wave, which is transmitted at the boundary between the air and thin films, then *reflected* from the substrate and transmitted at the boundary between the thin film and air, can be written by imposing the coefficients corresponding to each stage as:

$$E_2 = t_{01} t_{10} r_{12} \exp(-i\alpha) E_i \tag{17}$$

The higher-ordered reflected wave (*i.e.*, how many times the wave undergoes internal reflection within the thin film) can be given in the same way. We thus write the reflected wave considering the optical interference by the superposition of all reflected waves:

$$E_{012} = [r_{01} + t_{01} t_{10} \sum_{m=1}^{\infty} r_{10}^{m-1} r_{12}^m \exp(-mi\alpha)] E_i = r_{012} E_i \tag{18}$$

$r_{012}$ thus represents the reflection coefficient considering the optical interference. Using the formula for the sum of infinite geometric series, we can simply write Eq. 18 as:

$$r_{012} = r_{01} + \frac{t_{01} t_{10} r_{12} \exp(-i\alpha)}{1 - r_{10} r_{12} \exp(-i\alpha)}. \tag{19}$$

Using $r_{10} = -r_{01}$ and $t_{01} t_{10} = 1 - r_{01}^2$ obtained by Eqs. 3.1, 3.2, 5.1, and 5.2, $r_{012}$ can be further reduced[3]:

$$r_{012} = \frac{r_{01} + r_{12} \exp(-i\alpha)}{1 + r_{01} r_{12} \exp(-i\alpha)} \tag{20}$$

Using Eq. 6.1 and Eq. 20, we can obtain the reflectance of multilayered structures for a specific single wavelength. To simulate the color, we need to have the reflectance spectrum for the visible range, which thus requires different refractive indices and extinction coefficients depending on the wavelength (*i.e.*, dispersion).

*Color simulation*

To convert a reflectance spectrum to color (*i.e.*, sRGB values), the spectrum is first mapped to CIE 1931 XYZ color space using the formulas based on color matching functions[4]:

$$N = \int \bar{y}(\lambda) I(\lambda) \, d\lambda \tag{21.1}$$

$$X = \frac{1}{N} \int \bar{x}(\lambda) S(\lambda) I(\lambda) \, d\lambda \tag{21.2}$$

$$Y = \frac{1}{N} \int \bar{y}(\lambda) S(\lambda) I(\lambda) \, d\lambda \tag{21.3}$$

$$Z = \frac{1}{N} \int \bar{z}(\lambda) S(\lambda) I(\lambda) \, d\lambda \tag{21.4}$$

where $\lambda$ is the wavelength, and $\bar{x}$, $\bar{y}$, and $\bar{z}$ are the color-matching functions, which are the numerical description of the chromatic and tristimulus response of cone cells[5]. The parameters $S(\lambda)$ and $I(\lambda)$ are the spectral reflectance and standard illuminant, respectively. In this study, we employed CIE 1931 2° standard observer and D65 white reference as the color matching functions and the reference illuminant, respectively. The values $X$, $Y$, and $Z$ can be converted to either CIE xyY values or sRGB values. For CIE xyY values,

$$x = \frac{X}{X + Y + Z} \tag{22.1}$$

$$y = \frac{Y}{X + Y + Z} \tag{22.2}$$

here $x$ and $y$ are the coordinates of CIE 1931 color space chromaticity diagram, which are widely used to specify colors in practice. CIE XYZ coordinates are linearly transformed to sRGB coordinates using the following conversion matrix:

$$\begin{bmatrix} r \\ g \\ b \end{bmatrix} = \begin{bmatrix} 3.24045 & -1.53714 & -0.49853 \\ -0.96927 & 1.87601 & 0.04156 \\ 0.05564 & -0.20403 & 1.05723 \end{bmatrix} \begin{bmatrix} X \\ Y \\ Z \end{bmatrix} \tag{23}$$

where $r$, $g$, and $b$ are linear RGB. To obtain the sRGB, we use:

$$V = \begin{cases} 12.92v & v \leq 0.0031308 \\ 1.055v^{1/2.4} - 0.055 & \text{otherwise} \end{cases} \qquad (24)$$

where $v \in \{r, g, b\}$ and $V \in \{R, G, B\}$. Since the values obtained using Eq. 24 are in the nominal range (0 – 1), we multiply each component by 255 to represent conventional sRGB values in the range of 0 – 255.

# Python coding

Based on the aforementioned theoretical background, we developed a Python code to simulate the colors of multilayered thin films. We import libraries – NumPy, pandas, SciPy, and matplotlib for numerical computations, data manipulation, interpolation, and visualization, respectively. A wavelength array in the visible range with a fixed step size is generated using the *arange* function in the NumPy library (line 2 in Code 1). To apply the dispersion relation of each layer, we need to import the refractive indices ($n$) and extinction coefficients ($k$) for the wavelengths in the array (lines 3 – 14 in Code 1). In this study, we aim to simulate the color of $SnO_2$ layers on a Si substrate, which has a native oxide layer; thus we use the $n$ and $k$ values of $SnO_2$, $SiO_2$, and Si from the literature[6-9]. We import a standard illuminant (D65) and apply it to the obtained reflectance spectra to simulate the spectra from the sample, and the color-matching functions (CMFs) are imported and applied to extract RGB values from the simulated spectra (lines 15 – 22 in Code 1). Here, missing $n$, $k$, and CMFs values, which are required but absent in the literature, can be estimated by interpolation using *interp1d* function in SciPy library (lines 24 – 38 in Code 1). Since both the extinction coefficients and dispersion of air in the visible range are negligible, we set the refractive index of air as the constant: 1.0003 and define the imaginary unit as *j* (lines 45 – 48 in Code 1).

**Code 1.** Python code for data preprocessing.

```python
1  # Standard wavelength
2  wl = np.arange(380, 750.1, 1)
3  # Tin dioxide refractive index
4  sno2_n = pd.read_csv('SnO2 n.csv')
5  wl_sno2_n= sno2_n.iloc[:, 0].values; n_sno2 = sno2_n.iloc[:, 1].values
6  # Tin dioxide extinction coefficient
7  sno2_k = pd.read_csv('SnO2 k.csv')
8  wl_sno2_k = sno2_k.iloc[:, 0].values; k_sno2 = sno2_k.iloc[:, 1].values
9  # Silicon dioxide refractive index
10 sio2_n = pd.read_csv('SiO2 n.csv')
11 wl_sio2_n = sio2_n.iloc[:, 0].values; n_sio2 = sio2_n.iloc[:, 1].values
12 # Silicon refractive index and extinction coefficient
13 si = pd.read_csv('Si n k.csv')
14 wl_si = si.iloc[:, 0].values; n_si = si.iloc[:, 1].values; k_si = si.iloc[:, 2].values
15 # http://www.mtheiss.com/docs/code/?bk_d65_spectrum.htm
16 D65_spec = pd.read_csv("D65_step_1.csv")
17 D65 = D65_spec.iloc[0:371, 1].values
18 # http://cvrl.ucl.ac.uk/
19 # CMFs -> CIE 1931 2-deg, XYZ CMFs
20 cie_data = pd.read_csv('CIE_cc_1931_2deg_cvrl.csv')
21 cie_wavelength = cie_data.iloc[:, 0]
22 CMFs_X_b = cie_data.iloc[:, 1]; CMFs_Y_b = cie_data.iloc[:, 2]; CMFs_Z_b = cie_data.iloc[:, 3]
23
24 # Interpolation - Standard wavelength
25 interp_sno2_n = interp1d(wl_sno2_n, n_sno2, kind = 'cubic')
26 interpolated_sno2_n = interp_sno2_n(wl)
27 interp_sno2_k = interp1d(wl_sno2_k, k_sno2, kind = 'cubic')
28 interpolated_sno2_k = interp_sno2_k(wl)
29 interp_sio2_n = interp1d(wl_sio2_n, n_sio2, kind = 'cubic')
30 interpolated_sio2_n = interp_sio2_n(wl)
31 interp_si_n = interp1d(wl_si, n_si, kind = 'cubic')
32 interpolated_si_n = interp_si_n(wl)
33 interp_si_k = interp1d(wl_si, k_si, kind = 'cubic')
34 interpolated_si_k = interp_si_k(wl)
35 interp_X = interp1d(cie_wavelength, CMFs_X_b, kind = 'cubic')
36 interp_Y = interp1d(cie_wavelength, CMFs_Y_b, kind = 'cubic')
37 interp_Z = interp1d(cie_wavelength, CMFs_Z_b, kind = 'cubic')
38 CMFs_X = interp_X(wl); CMFs_Y = interp_Y(wl); CMFs_Z = interp_Z(wl)
39
40 # Summation n and k
41 N_sno2 = np.array([complex(interpolated_sno2_n[i], -interpolated_sno2_k[i]) for i in range(len(interpolated_sno2_k))], dtype=complex)
42 N_sio2 = np.array([complex(interpolated_sio2_n[i], -0) for i in range(len(interpolated_sio2_n))], dtype=complex)
43 N_si = np.array([complex(interpolated_si_n[i], -interpolated_si_k[i]) for i in range(len(interpolated_si_k))], dtype=complex)
44
45 # Air n, k
46 N_air = complex(1.0003, -0)
47 # imagenary unit
48 j = complex(0, 1)
```

The reflection coefficients of the p- and s-polarized waves and the reflectances are calculated in lines 1 – 10 and lines 11 – 45 at Code 2, respectively. Here, $N\_j$ and $N\_k$ indicate the complex refractive indices of media j and k, respectively. ang_j and ang_k are the angle of incidence and transmission, respectively. In line 11, wl, $N\_0$, $N\_1$, $N\_2$, and $N\_3$ are the wavelength, the complex refractive indices of the atmosphere (*i.e.*, air), top, middle, and bottom layers (here the bottom layer corresponds to the substrate), respectively. $d\_1$, and $d\_2$ are the thicknesses of the top and middle layers, respectively. The mathematical expression corresponding to each code is given above each code by the equation numbers described in Theoretical background.

**Code 2**. Python code to calculate the reflection coefficients and the reflectance.

```python
1   # Eq. (3.1)
2   def reflection_p(N_j, N_k, ang_j, ang_k):
3       r = ((N_k * np.cos(ang_j) - N_j * np.cos(ang_k))
4            / (N_k * np.cos(ang_j) + N_j * np.cos(ang_k)))
5       return r
6   # Eq. (5.1)
7   def reflection_s(N_j, N_k, ang_j, ang_k):
8       r = ((N_j * np.cos(ang_j) - N_k * np.cos(ang_k))
9            / (N_j * np.cos(ang_j) + N_k * np.cos(ang_k)))
10      return r
11  def reflectance(wl, N_0, N_1, N_2, N_3, d_1, d_2, angle):
12      # Snell's law
13      ang_0 = angle * np.pi / 180
14      ang_1 = np.arcsin(N_0/N_1 * np.sin(ang_0))
15      ang_2 = np.arcsin(N_0/N_2 * np.sin(ang_0))
16      ang_3 = np.arcsin(N_0/N_3 * np.sin(ang_0))
17      # Eq. (15)
18      alpha_1 = 4 * np.pi * d_1 * N_1 * np.cos(ang_1) / wl
19      alpha_2 = 4 * np.pi * d_2 * N_2 * np.cos(ang_2) / wl
20      # Reflection coefficient of p-polarized wave at each interface
21      r_01_p = reflection_p(N_0, N_1, ang_0, ang_1)
22      r_12_p = reflection_p(N_1, N_2, ang_1, ang_2)
23      r_23_p = reflection_p(N_2, N_3, ang_2, ang_3)
24      # Eq. (20)
25      r_123_p = ((r_12_p + r_23_p * np.exp(-j * alpha_2))
26                 / (1 + r_12_p * r_23_p * np.exp(-j * alpha_2)))
27      # Eq. (20)
28      r_0123_p = ((r_01_p + r_123_p * np.exp(-j * alpha_1))
29                  / (1 + r_01_p * r_123_p * np.exp(-j * alpha_1)))
30      # Reflection coefficient of s-polarized wave at each interface
31      r_01_s = reflection_s(N_0, N_1, ang_0, ang_1)
32      r_12_s = reflection_s(N_1, N_2, ang_1, ang_2)
33      r_23_s = reflection_s(N_2, N_3, ang_2, ang_3)
34      # Eq. (20)
35      r_123_s = ((r_12_s + r_23_s * np.exp(-j * alpha_2))
36                 / (1 + r_12_s * r_23_s * np.exp(-j * alpha_2)))
37      # Eq. (20)
38      r_0123_s = ((r_01_s + r_123_s * np.exp(-j * alpha_1))
39                  / (1 + r_01_s * r_123_s * np.exp(-j * alpha_1)))
40      # Eq. (6.1)
41      R_p = np.abs(r_0123_p)**2
42      R_s = np.abs(r_0123_s)**2
43      # Eq. (7)
44      R_n = (R_p+R_s)/2
45      return R_n
```

By modifying the values of d_1, d_2, and angle in line 11 of Code 2, we can obtain reflectance spectra for various thicknesses and angles of incidence. To convert the reflectance spectra into RGB values, we apply Code 3; in line 1, R_n is the calculated reflectance spectrum, and D65, CMFs_X, CMFs_Y, and CMFs_Z are the standard illuminant, and the color matching functions for X, Y, and Z, respectively, which are imported and defined earlier (Code 1).

**Code 3**. Python code to convert the reflectance spectrum to RGB values.

```
1   def reflectance2RGB(R_n, CMFs_X, CMFs_Y, CMFs_Z, D65):
2       # Reflectance to CIE XYZ
3       # http://www.brucelindbloom.com/index.html?Eqn_XYZ_to_RGB.html
4       # Eq. (21.1)
5       N = sum(CMFs_Y * D65)
6       # Eq. (21.2)
7       X = sum(CMFs_X * D65 * R_n) / N
8       # Eq. (21.3)
9       Y = sum(CMFs_Y * D65 * R_n) / N
10      # Eq. (21.4)
11      Z = sum(CMFs_Z * D65 * R_n) / N
12      XYZ = np.array([X, Y, Z])
13      # RGB working space: sRGB / reference white D65
14      # XYZ to RGB matrix
15      # http://www.brucelindbloom.com/index.html?Eqn_XYZ_to_RGB.html
16      m = np.array([[3.24045, -1.53714, -0.49853],
17                    [-0.96927, 1.87601, 0.04156],
18                    [0.05564, -0.20403, 1.05723]])
19      # Eq. (23)
20      rgb_linear = np.dot(m, XYZ)
21      RGB = np.zeros(3)
22      # Eq. (24)
23      if (rgb_linear[0] <= 0.0031308):
24          RGB[0] = 12.92 * rgb_linear[0]
25      else:
26          RGB[0] = 1.055 * np.abs(rgb_linear[0]) ** (1/2.4) - 0.055
27      if (rgb_linear[1] <= 0.0031308):
28          RGB[1] = 12.92 * rgb_linear[1]
29      else:
30          RGB[1] = 1.055 * np.abs(rgb_linear[1]) ** (1/2.4) - 0.055
31      if (rgb_linear[2] <= 0.0031308):
32          RGB[2] = 12.92 * rgb_linear[2]
33      else:
34          RGB[2] = 1.055 * np.abs(rgb_linear[2]) ** (1/2.4) - 0.055
35      RGB_res[i] = np.round(np.clip(RGB * 255, 0, 255))
```

**Results**

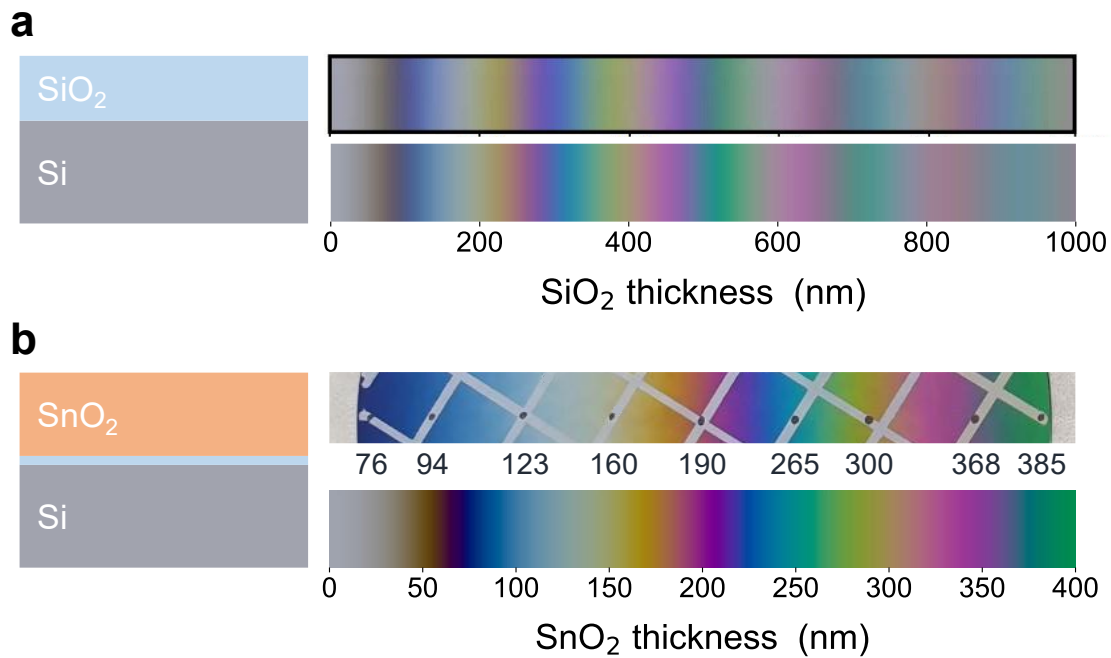

**Figure 4.** Comparison of simulated colors with literature and experiment: (a) SiO$_2$ thin film on a Si substrate: simulated colors as a function of the SiO$_2$ film thickness from literature (upper)[10] and by Python code in this study (lower). (b) SnO$_2$ thin film on a Si substrate: the photograph of the SnO$_2$ thin film with thickness gradient (upper) and simulated color as a function of the SnO$_2$ film thickness by the Python code (lower). The grid in the photograph is a masked area for line profile measurements – the value below each dot is the film thickness from the line profile measurements nearby. The thickness of a native oxide layer was set to 2 nm in the simulation.

To validate the Python code for simulating the color of multilayers, we simulate the colors of two different oxide thin films (SiO$_2$ and SnO$_2$) on a Si substrate as a function of the film thickness, and compare them with literature and experiment, respectively. We first checked the color of SiO$_2$ thin film on a Si substrate as a function of the film thickness. It can be clearly seen that the obtained color chart is very consistent with the reported simulation result (Figure 4a). For the comparative analysis with experiment, we fabricated the SnO$_2$ film with thickness gradient on a Si wafer using a sputtering method and measured thicknesses by line profile measurement across a masked area using noncontact (tapping mode) atomic force microscopy (AFM; Icon-PT-PLUS, Bruker). The base pressure was ≈ 3 × 10$^{-8}$ Torr, and the working pressure of 10 mTorr was adjusted by Ar (99.9999%). The films were grown at room temperature. We assumed the presence of a 2 nm thick native oxide layer (SiO$_2$) between the film and substrate[11,12]. The measured thickness and the apparent color nearby coincide with the simulated color chart as a function of thickness (Figure 4b). These results attest to the validity of the Python code and suggest that this code can be further developed to estimate the thickness of a film with known *n* and *k* values and vice versa.

**Conclusion**

We demonstrated the Python script for simulating the color of multilayered structures, with detailed theoretical background and annotations for readers as well as users. We believe that anyone can adjust structural and material parameters for their own system and research by following the instructions presented here. It is also worthwhile noting that the code generates a substantial quantity of data very rapidly – for example, simulating 15000 colors represented took merely 12 seconds (using CPU: Intel i7-1255U and GPU: Intel Iris Xe Graphics). This code can be further incorporated and developed for various purposes – especially for machine learning, which requires many spectral and color data for model training.

**Acknowledgement**

This work was supported by the National Research Foundation of Korea Grant funded by the Korean Government (NRF-2021R1C1C1009863) and Regional Innovation Strategy (RIS) through the National Research Foundation of Korea (NRF) funded by the Ministry of Education (MOE) (2023RIS-007).